# Comment on the "Janus Point" Explanation of the Arrow of Time

H. Dieter Zeh – www.zeh-hd.de - Jan. 2016

Julian Barbour and his collaborators have recently claimed to be able to explain the arrow of time in terms of their shape dynamics by means of a "Janus point" solution that would not require an improbable initial condition of very low entropy.[1,2,3] Although this proposal has aroused several positive reactions in the secondary media, it appears trivial when considered in conventional terms, and far from being novel or realistic.

If an ensemble of unbound objects is "initially" placed in a partial (but physically open) volume element with arbitrary initial velocities of the objects, it will very soon fly apart in both directions of time, that is, in any time direction of calculation as a consequence of the time reversal symmetry of the dynamical laws. Thereby, the information representing its initial localization is dynamically transformed into information about thermodynamically "irrelevant" information about correlations between relative position and velocity. Quite generally, such an initial condition of low physical (coarse-grained) entropy allows the latter to increase in the direction of calculation, while ensemble entropy is conserved.[4] The double arrow is evidently a consequence of the (improbable) condition that was applied at the start of the calculation and the assumption that time may be represented by the real numbers. If both legs of the trajectory are thus assumed to describe reality, the initial condition for the calculation becomes a middle condition (a "Janus point") for the formal time evolution according to growing time parameter *t*. This entropy minimum is, therefore, well known in Statistical Mechanics to appear under such a procedure (and sometimes regarded as a paradox because it is never observed). A generic distribution of a finite number of objects in infinite space would instead *never* be found in a finite volume, and the numerically studied examples are *not at all* "typical" – as claimed. (Anyhow, what is a typical universe if we have just one?) Even if there were some encounters between these objects under somewhat relaxed conditions, they could as well happen at quite different times and places. However, *if* all other arrows could be derived from an arrow of expansion of the universe (as has often been hypothesized for a big bang model), a physical observer would indeed always experience the direction of expansion as his future, which he cannot "consistently remember" – in contrast to part of his past.

If the philosophy of shape dynamics requires that scale (hence volume) be physically meaningless, the arguments simply demonstrate that this philosophy is wrong at this level of description of reality. Scale is incorporated into classical phase space volume, that can be absolutely measured in terms of Planck's constant, for example. A *change* in time of the measure of scale would even be incompatible with Newton's laws (and thus cannot be meaningless), while the hypothetical scale invariance of an elusive *fundamental* theory of everything would have to presume time-asymmetric and non-unitary quantum phenomena, such as symmetry breaking and decoherence, as an important *prerequisite* for describing the observed quasi-classical world with its effective scales.[5] For example, these processes may give rise to the birth of new effective degrees of freedom with their new entropy capacity.[6] Fundamental problems (including the "problem of time") can hardly be solved or even be formulated in classical terms.[7,8] This specific quantum aspect of timelessness may not be made particularly clear in Barbour's very popular book[9] and some subsequent publications.

If the classical "objects" that form the model discussed by the authors are assumed to interact by means of attractive forces, one has to require non-negative total energy (such as zero energy in accordance with shape dynamics) in order to obtain similar conclusions as just described for free objects, while some clusters of these objects may then form and remain in bound states. For long-range forces, such as gravity, the evolution along each direction of calculation may also lead to the irreversible formation of large-scale inhomogeneities and structure in this case. This irreversibility is a consequence of an initially assumed approximate homogeneity of the spatial distribution (far from forming black holes), which is again an improbable condition under gravity. The evolution within the considered Newtonian model may then possibly be characterized by the authors' novel measure of complexity as an elementary concept of "order", although this remains to be shown in detail. In General Relativity, this process would lead to black holes by means of the evaporation of some objects from their clusters and, more realistically, by radiation of any kind (when *presuming* a radiation arrow in the form of retardation, that is, the further improbable condition that radiation be negligible at the Janus point). Both kinds of energy loss lead to *heating* of the clusters and to entropy generation because of the negative heat capacity of gravitating systems.[10] This consequence is particularly important for the arising thermal non-equilibrium between hot stars and cold space that is known to lead to more complex (more realistic) forms of order by self-organization under appropriate interactions.

The unreasonable condition of no more than a finite number of objects, located in a finite element of infinite space, seems to be irrelevant for this irreversible gravitational contraction (in contrast to the condition of initial homogeneity). Therefore, the *essential* difference between the models discussed by Barbour, Koslowski and Mercati and conventional models of an expanding and gravitating universe is just the questionable exclusion of a space-time singularity at highest matter density, such as a big bang, while the starting condition of very low physical entropy for the calculation perfectly describes the arrow of time at least as far as the mentioned fundamental *quantum* phenomena are neglected (which seems to be unrealistic even today, however). These phenomena require either an asymmetric quantum dynamical law, such as a fundamental collapse, or – in order to facilitate an irreversible decoherence process – the (again improbable) initial or middle condition of negligible nonlocal entanglement, in analogy to the retarded nature of all *statistical* correlations that makes them "irrelevant" for the future and is also known as Boltzmann's chaos assumption. The relation between this retardation of correlations and the initial homogeneity is as yet not fully understood.[11]

---